\documentclass[a4paper,aps,nofootinbib]{revtex4}

\usepackage{amsmath}
\usepackage{amssymb}

\def\beq{\begin{equation}}
\def\eeq{\end{equation}}

\def\beqn{\begin{eqnarray}}
\def\eeqn{\end{eqnarray}}

\begin{document}

\title{Dirac-K\"ahler Theory and Massless Fields}

\author{V. A. Pletyukhov}
\affiliation{Brest State University, Brest, Belarus}
\author{V. I. Strazhev}
\affiliation{Belarusian State University, Minsk, Belarus}

\begin{abstract}
 Three massless limits of the Dirac-K\"ahler theory are considered.
 It is shown that the Dirac-K\"ahler equation for massive particles
 can be represented as a result of the gauge-invariant mixture
 (topological interaction) of the above massless fields.

\end{abstract}

\maketitle

\section{Introduction}
In the 1980's, it was shown in [1] with the use of differential
forms apparatus that the the Dirac-K\"ahler equation (DKE)
is correct in describing Dirac particles (quarks) in the
lattice formulation of QCD. After that the DKE has become
to attract attention of many theorists. Note that
the name "Dirac-K\"ahler equation" was introduced in [1]
although the vector form of the DKE was discovered by Darwin [2].
His aim was to find an equation of motion of an electron that would
be equivalent to the Dirac equation but without using spinors.
Fundamental properties of the DKE were established by K\"ahler [3].
Later on, the DKE has been rediscovered in different mathematical
formulations (see, e.g., ref. [4, pp.38,51] and references therein).

One should emphasize that up to now only massive DKE have been studied.
The massless limit of the DKE has not been investigated in detail yet.
The present paper is aimed at making up such a deficiency.

\section{The Dirac-K\"ahler equation}
The DKE is equivalent to the following tensor system
\begin{subequations}
\label{sys1}
\beqn
\label{sys1a}
\partial_\mu \psi_\mu + m\psi_0  =0,
\\
\label{sys1b}
\partial_\mu\tilde {\psi }_\mu + m\tilde {\psi }_0 =0,
\\
\label{sys1c}
\partial_\nu \psi_{\left[ {\mu \nu } \right]}
+ \partial_\mu \psi _0 + m\psi_\mu = 0,
\\
\label{sys1d}
\frac{1}{2}\varepsilon_{\mu\nu\alpha\beta}
\partial_\nu \psi_{\left[ {\alpha\beta } \right]}
+\partial_\mu \tilde {\psi }_0 +m\tilde{\psi }_\mu = 0,
\\
 - \partial_\mu \psi_\nu + \partial_\nu \psi_\mu  +
\varepsilon_{\mu\nu \alpha \beta } \partial_\alpha
\tilde {\psi }_\beta  +
m\psi _{\left[{\mu \nu } \right]}  = 0,
\label{sys1e}
\eeqn
\end{subequations}
where $\tilde {\psi }_\mu =\frac{1}{3!}~\varepsilon _{\mu \nu
\alpha \beta } \psi_{\left[ {\nu \,\alpha \beta } \right]}$ is an axial vector,
$\tilde {\psi }_0  =
\frac{1}{4!}~\varepsilon_{\mu \nu \alpha \beta }
\psi_{\left[ {\mu \nu \alpha \beta } \right]} $ is a pseudoscalar, and
$\varepsilon_{\mu \nu \alpha \beta } $ is the Levi-Civita tensor
($\varepsilon _{1234}=- i)$.

The set of equations (\ref{sys1}) can be brought to
the following form
\beqn
\label{2.2}
\left( {\Gamma _\mu \partial_\mu + m} \right)~\Psi  = 0,
\eeqn
where $\Psi $ is the 16-component wave function
\begin{equation}
\label{Psi}
\Psi \equiv \Psi_A :\psi _0 ,\tilde {\psi }_0, \psi _\mu ,\tilde{\psi }_\mu ,
\psi_{\left[ {\mu\nu } \right]} ,
\end{equation}
consisting of the Dirac-K\"ahler (DK) field components which
 form the full set of antisymmetric
tensor fields in the space of dimension $d=4$. The $16\times 16$
matrices $\Gamma_\mu$ satisfy the anticommutation rules analogous to these
for the Dirac matrices
\begin{equation}
\label{eq8}
\Gamma_\mu \Gamma_\nu  + \Gamma_\nu \Gamma_\mu =2~\delta_{\mu\nu }.
\end{equation}

In respect to the theory of relativistic wave equations, the system of the DKE
describes a particle with a single mass $m$, variable spin 0 or 1, and
with double degeneration of states over an additional
quantum number (internal parity). At the same time, the Lagrangian of the DK field
is invariant under a transformation of the group of internal (dial) symmetry
$SO\,\left( {4,\,2\,} \right)$ [4, pp.28,35]. Group generators have the
following form
\beq
{\Gamma }'_\mu \,,\,\,\,\,{\Gamma }'_{\left[ \mu \right.} \,{\Gamma
}'_{\left. \nu \right]} \,,\,\,\,\,\,{\Gamma }'_5 \,{\Gamma }'_\mu
\,,\,\,\,\,{\Gamma }'_5 \,\,.
\eeq
Here  ${\Gamma }'_\mu $ is second set of $16\times 16$ matrices satisfying
the Dirac matrix algebra and commuting with ${\Gamma }_\mu $.
The above properties of the symmetry are easily checked if one takes
into account that in the so-called fermion basis
(see, e.g., the book [4, p.72])
the matrices $\Gamma_\mu$ and $\Gamma '_\mu$ can be written as
\begin{equation}
\label{eq11}
\Gamma _\mu  = I_4 \,\, \otimes \,\,\gamma _\mu
\,\,,\,\,\,\,\,\,\,\,{\Gamma }'_\mu =\gamma _\mu \,\, \otimes
\,\,I_4,
\end{equation}
where $\gamma_\mu$ are the Dirac matrices, $I_4$ is the unit 4 by 4 matrix.

Internal ("dial") symmetry transformations relate with each other
tensors of different ranks.
Thereby, the theoretical group ground is established for association of
a Dirac particle with the DK field.
This particle (''geometric fermion'') in addition to the spin $\frac 12$
has its inner degrees of freedom which are of space-time origin.
For instance, if one turns on electromagnetic interaction in the standard
way, i.e. by the replacement $\partial _\mu \to \partial_\mu  - ieA_\mu $
($A_\mu$ is an electromagnetic potential) then the DKE will have
solutions equivalent to these for the Dirac equation
because the matrices in both equations have the same algebraic properties.

Since we are going to study massless limits of the DKE it is convenient
to transform the system (\ref{sys1}) to another form
replacing the common mass $m$ by two new mass parameters, namely
a parameter $m_1$ in (\ref{sys1a}), (\ref{sys1b}) and (\ref{sys1e}),  and
a parameter $m_2$ in (\ref{sys1c}) and (\ref{sys1d}).
One has the new system
\begin{subequations}
\label{sys5}
\beqn
\label{sys1a1}
\partial_\mu \psi_\mu + m_1\psi_0  =0,
\\
\label{sys1b1}
\partial_\mu\tilde {\psi }_\mu + m_1\tilde {\psi }_0 =0,
\\
\label{sys1c1}
\partial_\nu \psi_{\left[ {\mu \nu } \right]} +
\partial_\mu \psi _0 + m_2\psi_\mu = 0,
\\
\label{sys1d1}
\frac{1}{2}\varepsilon_{\mu\nu\alpha\beta}
\partial_\nu \psi_{\left[ {\alpha\beta } \right]}
+\partial_\mu \tilde {\psi }_0 +m_2\tilde{\psi }_\mu = 0,
\\
 - \partial_\mu \psi_\nu + \partial_\nu \psi_\mu  +
\varepsilon_{\mu\nu \alpha \beta } \partial_\alpha
\tilde {\psi }_\beta  +
m_1\psi _{\left[{\mu \nu } \right]}  = 0.
\label{sys1e1}
\eeqn
\end{subequations}
Matrix form of this system is as follows
\begin{equation}
\label{newsyst}
\left( {\Gamma _\mu \,\partial _\mu \,\, + \,\,m_1 \,P_1 \,\, + \,\,m_2
\,P_2 } \right)\,\Psi =0,
\end{equation}
where $P_1$ and $P_2$ are the projection operators with the properties
\begin{equation}
\label{eq13}
\begin{array}{l}
 P_1^2  = P_1, \quad P_2^2  = P_2, \quad  P_1+P_2 = 1,\quad P_1P_2=0,
 \\
 P_1\Gamma_\mu +\Gamma_\mu P_1=\Gamma_\mu, \quad
P_2\Gamma_\mu +\Gamma_\mu P_2=\Gamma_\mu.
 \end{array}
\end{equation}

A second order equation equivalent to the system (\ref{sys5}) is
 \begin{equation}
\label{eq15}
\left(\Box - m_1 m_2 \right)~\psi _A=0.
\end{equation}
This means that the system  (\ref{sys5})
at $m_1\neq 0$ and $m_2\neq 0$ describes a particle with a single
mass $m=\sqrt{m_1m_2}$, i.e. it does not differ from the usual DKE.
But if any of the mass parameters (or both simultaneously) is equal to zero
then the system (\ref{sys5}) and the equation
(\ref{newsyst}) correspond to the massless case. We proceed now to
the study of this important case.

\section{ "Electromagnetic" massless limit  }

Let us suppose in (\ref{sys5}) and (\ref{newsyst}) that
\beq
m_2=0.
\eeq
If $m_1\neq 0$, then without loosing of generality  one can put $m_1=1$
so that the system (\ref{sys5}) is now
\begin{subequations}
\label{sys3}
\beqn
\label{sys1a2}
\partial_\mu \psi_\mu + \psi_0  =0,
\\
\label{sys1b2}
\partial_\mu\tilde {\psi }_\mu + \tilde {\psi }_0 =0,
\\
\label{sys1c2}
\partial_\nu \psi_{\left[ {\mu \nu } \right]} +
\partial_\mu \psi _0 = 0,
\\
\label{sys1d2}
\frac{1}{2}\varepsilon_{\mu\nu\alpha\beta}
\partial_\nu \psi_{\left[ {\alpha\beta } \right]}
+\partial_\mu \tilde {\psi }_0 = 0,
\\
 - \partial_\mu \psi_\nu + \partial_\nu \psi_\mu  +
\varepsilon_{\mu\nu \alpha \beta } \partial_\alpha
\tilde {\psi }_\beta  +  \psi _{\left[{\mu \nu } \right]}  = 0,
\label{sys1e2}
\eeqn
\end{subequations}
and correspondingly
\begin{equation}
\label{P_1}
\left( {\Gamma _\mu \partial _\mu  + P_1 } \right)\,\Psi
=0.
\end{equation}
Note that the transition to the massless limit by means of
introduction of projective operators into (\ref{2.2}) is
a typical procedure for fields with integer spins. For example,
it is carried out when one wants to pass on to massless vector field
(i.e. electromagnetic field) from the massive Duffin-Kemmer equation.

In order to establish the physical meaning of the system
(\ref{sys3}), we note, first of all, that the field
functions $\Psi_A$ obey the D'Alembert equation
\beq
\label{Dalam}
\Box \Psi_A=0,
\eeq
i.e. this system describes a massless field.
The vector $\psi_\mu(x) $ and the pseudovector $\tilde \psi_\mu(x)$ play role of
potentials in (\ref{sys3}), and the antisymmetric tensor
$\psi_{{\left[ {\mu \,\nu}  \right]}} $ is an intensity tensor.  Physical meaning
of the scalar, $\psi_0(x), $ and pseudoscalar, $\tilde \psi_0(x)$, functions
will be clarified later on.

The system (\ref{sys3}) is invariant under gauge
transformations of the potentials
\begin{equation}
\label{gaugetr}
\delta \psi_\mu \left( x \right)=\partial_\mu\lambda(x),\quad
\delta \tilde {\psi }_\mu (x)=\partial_\mu \tilde {\lambda }(x),
\end{equation}
where $\lambda(x)$ and $\tilde {\lambda }(x)$ satisfy the following conditions
\beq
\label{lambdas}
\Box ~\lambda(x)=0,\quad \Box~\tilde {\lambda}(x)=0.
\eeq

To find independent physical states of the field system under consideration,
we perform the Fourier transformation
\begin{subequations}
\label{foure}
\beqn
\label{foure1}
 \psi _\mu \,\left( x \right)&=&\int {\psi _\mu \,\left(
{\underline{p}} \right)\,e^{i\,p\,x}\,d^3p}+ h.c.,
\label{foure2}
\\
\tilde {\psi }_\mu \,\left( x \right)&=&
\int {\tilde {\psi}_\mu \,\left( {\underline{p}} \right)\,e^{i\,p\,x}\,d^3p}+ h.c.,
\\
\label{foure3}
\psi_0 \,\left( x \right)&=&
\int {\psi _0 \,\left({\underline{p}} \right)\,e^{i\,p\,x}\,d^3p}+ h.c.,
\\
\tilde \psi_0 \,\left( x \right)&=&
\int {\tilde \psi _0 \,\left({\underline{p}} \right)\,e^{i\,p\,x}\,d^3p}+ h.c.
\label{foure4}
\eeqn
\end{subequations}

Let us expand the amplitudes
$\psi _\mu \,\left( {\underline{p}} \right)$ and
$\tilde {\psi }_\mu \,\left( {\underline{p}} \right)$ over the complete basis
$e_\mu ^{\left( 1 \right)},\,e_\mu ^{\left( 2 \right)} ,\,p_\mu, \,n_\mu $ with properties \cite{ogievetsky66}
\beqn
\label{basis}
e_\mu ^{\left( i \right)} e_\mu^{\left( j \right)} =\delta_{i\,j},~
e_\mu ^{\left( i \right)} p_\mu=0,~
e_\mu ^{\left( i \right)} n_\mu =0,
\nonumber
\\
p_\mu ^2 = 0, ~       n_\mu^2=-1.
\eeqn
Note that the basis (\ref{basis}) is not orthogonal because it
contains an isotropic vector $p_\mu$. The desired decompositions can be
written as
\begin{equation}
\label{eq27}
\begin{array}{l}
 \psi _\mu \,\left( {\underline{p}} \right)\,\,\, = \,\,\,\sum\limits_{i\, =
\,1}^2 {a_i \,e_\mu ^{\left( i \right)} } \,\, + \,\,b\,p_\mu \,\, +
\,\,c\,n_\mu \,\,, \\
 \tilde {\psi }_\mu \,\left( {\underline{p}} \right)\,\,\, =
\,\,\,\sum\limits_{i\, = \,1}^2 {\tilde {a}_i \,e_\mu ^{\left( i \right)} }
\,\, + \,\,\tilde {b}\,p_\mu \,\, + \,\,\tilde {c}\,n_\mu.
 \end{array}
\end{equation}

Now let us take into account that due to (\ref{lambdas})
the gauge functions $\lambda (x)$ and $\tilde {\lambda }(x)$ have
the form analogous to (\ref{foure3}) and (\ref{foure4})
\begin{equation}
\label{eq28}
\begin{array}{l}
 \lambda \,\left( x \right)\,\,\, = \,\,\,\int {\lambda \,\left(
{\underline{p}} \right)\,e^{i\,p\,x}\,d^3p}+h.c.,
\\
\\
 \tilde {\lambda }\,\left( x \right)\,\,\, = \,\,\,\int {\tilde {\lambda
}\,\left( {\underline{p}} \right)\,e^{i\,p\,x}\,d^3p} + h.c,
 \end{array}
\end{equation}
where $\lambda \left( {\underline{p}} \right)$ and $ \tilde {\lambda
}\,\left( {\underline{p}} \right)$ are arbitrary amplitudes. Inserting
decompositions (\ref{foure1})--(\ref{foure4}) in (\ref{gaugetr}) we obtain
gauge transformations for the amplitudes of the potentials
\begin{equation}
\label{eq29}
\begin{array}{l}
\delta \,\psi _\mu \,\left( {\underline{p}} \right) = i\lambda
\,\left( {\underline{p}} \right)\,p_\mu,
\\
\\
\delta \,\tilde
{\psi }_\mu \,\left( {\underline{p}} \right) = i\,\tilde
{\lambda }\,\left( {\underline{p}} \right)\,p_\mu ,
\end{array}
\end{equation}
which mean that the amplitudes  $\psi _\mu( {\underline{p}})$ and
$\tilde {\psi }_\mu \,\left( {\underline{p}} \right)$
are defined up to unessential terms $i\,\lambda \,\left(
{\underline{p}} \right)\,p_\mu $ and $i\,\tilde {\lambda }\,\left(
{\underline{p}} \right)\,p_\mu $, respectively. Role of such terms in
the decompositions (\ref{eq27}) play  $b\,p_\mu $ and $\tilde
{b}\,p_\mu $. Omitting them we obtain for the amplitudes
$\psi _\mu (\underline{p})$ and $\tilde {\psi }_\mu \,\left( {\underline{p}}
\right)$
\begin{equation}
\label{eq30}
\begin{array}{l}
 \psi _\mu \,\left( {\underline{p}} \right) = \sum\limits_{i\, =
\,1}^2 {a_i \,e_\mu ^{\left( i \right)} } \,\, + \,\,c\,n_\mu \,\,, \\
 \tilde {\psi }_\mu \,\left( {\underline{p}} \right) =
\sum\limits_{i\, = \,1}^2 {\tilde {a}_i \,e_\mu ^{\left( i \right)} }
\,\, + \,\,\tilde {c}\,n_\mu.
 \end{array}
\end{equation}
Note that longitudinal oscillations (degrees of freedom) are absent
in (\ref{eq30}).

Scalar degrees of freedom are eliminated at the second quantization procedure
when the equations (\ref{sys1a2}) and (\ref{sys1b2}) for quantized field
are formulated in the form of conditions imposed on wave functions
 $\Psi _{phys} $ in the state space
\begin{equation}
\label{eq31}
\begin{array}{l}
\left( \partial_\mu \hat \psi_\mu (x)+\hat {\psi }_0 (x) \right)_+ \Psi _{phys}=0,
\\
\\
\left( \partial_\mu \hat {\tilde \psi}_\mu (x)+\hat {\tilde\psi }_0 (x) \right)_
+ \Psi _{phys}=0,
 \end{array}
\end{equation}
where the index $"+"$ means that the corresponding operator
contains the positive-frequency part only.
Keeping in mind the relations  (\ref{foure}), (\ref{basis}) and (\ref{eq30})
we obtain from (\ref{eq31})
\begin{equation}
\label{eq32}
\begin{array}{r}
 \left( \int \omega \left( \hat d-\hat c\right)
 e^{ipx}d^3p\right)\,\Psi _{phys}=0,
  \\
  \\
  \left( \int \omega \left( \hat {\tilde d}-\hat {\tilde c}\right)
 e^{ipx}d^3p\right)\,\Psi _{phys}=0,
  \end{array}
\end{equation}
where
\begin{equation}
\label{eq33}
d=\frac{\psi _0 \,\left( p \right)}{\omega
}\,\,,\,\,\,\,\,\,\,\,\tilde {d} = \frac{\tilde {\psi }_0
\,\left( p \right)}{\omega }
\end{equation}
play a role of the amplitudes of the scalar fields
 $\psi _0$ and $\tilde {\psi }_0 $.
It is follows from (\ref{eq32}) that for all $p$, the function $\Psi _{phys}$
has to satisfy the following conditions
\begin{equation}
\label{eq34}
\left( \hat d-\hat c\right)\,\Psi _{phys}=0,
  \\
\left( \hat {\tilde d}-\hat {\tilde c}\right)\Psi _{phys}=0,
\end{equation}

A standard procedure used to eliminate longitudinal and scalar oscillations
at quantization of electromagnetic field [6,pp.56,68] leads to relations
\cite{plet2008}
\begin{equation}
\label{eq35}
\begin{array}{r}
 \left( \Psi_{phys}, \left( \hat {d}^+\hat d +\hat {c}^+\hat c \right)
 \Psi_{phys}\right)=0,
 \\
 \\
\left( \Psi_{phys}, \left( \hat {\tilde{d}}^+~ \hat {\tilde d}
 +\hat {\tilde c}^+~\hat {\tilde c} \right)
 \Psi_{phys}\right)=0.
 \end{array}
\end{equation}

Due to (\ref{eq35}) the mean values disappear in the part
of the energy operator that contains scalar oscillations of both types.
Therefore, the system (\ref{sys3}) describes a massless vector field with
double degeneration of states. A special case of such a field is
the usual electromagnetic field.
As is follows from the previous analysis, there are no physical states
corresponding to the scalar and pseudoscalar functions
$\psi _0(x)$ and $\tilde {\psi }_0 (x)$.
These functions serve as gauge fields (``ghosts'').

\section{Notoph (Kalb-Ramond field)}

Let us consider the following case of the system (\ref{sys5}):
\begin{equation}
m_1=0, \quad m_2=1.
\end{equation}
In this case one has for (\ref{sys5})
\begin{subequations}
\label{sys4}
\beqn
\label{sys4a}
\partial _\mu \,\psi_\mu =0,
\\
\label{sys4b}
\partial _\mu \,\tilde {\psi }_\mu =0,
\\
\label{sys4c}
\partial _\nu \,\psi _{\left[ {\mu \,\nu } \right]}  + \partial _\mu
\,\psi _0  + \psi _\mu  = 0,
\\
\label{sys4d}
\frac{1}{2}\,\varepsilon _{\mu \,\nu \,\alpha \,\beta } \,\partial _\nu
\,\psi _{\left[ {\alpha \,\beta } \right]}  + \partial _\mu \,\tilde
{\psi }_0  + \tilde {\psi }_\mu  = 0,
\\
\label{sys4e}
 - \partial _\mu \,\psi _\nu  +\partial _\nu \,\psi _\mu  +
\varepsilon _{\mu \,\nu \,\alpha \,\beta } \,\partial _\alpha \,\tilde
{\psi }_\beta = 0.
\eeqn
\end{subequations}
Here $\psi _0$, $\tilde {\psi }_0$
and $\psi _{\left[ {\mu \,\nu } \right]} $ serve as potentials and
the vectors $\psi _\mu $ and $\tilde {\psi }_\mu $ serve as intensities.
The equations (\ref{sys4c}) and (\ref{sys4d}) are definitions of
intensities via potentials and
the equations (\ref{sys4a}), (\ref{sys4b}), and (\ref{sys4e})
are equations of motion.

The matrix form of  the system (\ref{sys4}) is
\begin{equation}
\label{eq42}
\left( {\Gamma _\mu \,\partial _\mu \,\, + \,\,P_2 } \right)\,\Psi  = 0.
\end{equation}
Using either tensor or matrix formulation of the field system under
consideration, one can easily show that all components of the wave function
obey the D'Alembert equation (\ref{Dalam}), i.e. again we deal with
the massless field.

Further discussion becomes to be more convenient if one introduces
an auxiliary tensor $\tilde \psi _{\left[ {\mu \,\nu } \right]} $
\begin{equation}
\label{eq43}
\tilde {\psi }\,_{\left[ {\mu \,\nu } \right]}  =
\frac{1}{2}\,\varepsilon _{\mu \,\nu \,\alpha \,\beta } \,\psi
\,_{\left[ {\alpha \,\beta } \right]} \,\,.
\end{equation}
One has for (\ref{sys4d})
\begin{equation}
\label{eq44}
\partial _\nu \,\tilde {\psi }\,_{\left[ {\mu \,\nu } \right]} \,\, +
\,\,\partial _\mu \,\tilde {\psi }_0 \,\, + \,\,\tilde {\psi }_\mu  =0.
\end{equation}
This equation will be used below together with (\ref{sys4d}).

In order to pass on to the momentum representation, we write down the
scalar potentials in form of (\ref{foure3}), (\ref{foure4})
and tensor potentials in the following form
\begin{equation}
\label{eq45}
\begin{array}{l}
 \psi \,_{\left[ {\mu \,\nu } \right]} \,\left( x \right)=
\int{\psi \,_{\left[ {\mu \,\nu } \right]} \,\left( {\underline{p}}
\right)\,e^{i\,p\,x}\,d^3\,p + } h.c.,
\\
\\
 \tilde {\psi }\,_{\left[ {\mu \,\nu } \right]} \,\left( x \right)=
\int {\tilde {\psi }\,_{\left[ {\mu \,\nu } \right]} \,\left(
{\underline{p}} \right)\,e^{i\,p\,x}\,d^3\,p + } h.c.
 \end{array}
\end{equation}

We expand now the amplitudes  $\psi \,_{\left[ {\mu \,\nu } \right]} \,\left(
{\underline{p}} \right)$ and $\tilde {\psi }\,_{\left[ {\mu \,\nu }
\right]} \,\left( {\underline{p}} \right)$ over the complete basis (\ref{basis})
\begin{subequations}
\beqn
 \psi \,_{\left[ {\mu \,\nu } \right]} \,\left( {\underline{p}}
\right)&=&f\,\left( {e_\mu ^{\left( 1 \right)} \,e_\nu ^{\left(
2 \right)} - e_\nu ^{\left( 1 \right)} \,e_\mu ^{\left( 2 \right)}
} \right)+
\nonumber
\\
&&+~\sum\limits_{i\, = \,1}^2 {g_i \,\left( {e_\mu
^{\left( i \right)} \,p_\nu \,\, - \,\,e_\nu ^{\left( i \right)} \,p_\mu }
\right)} +
\nonumber
\\
&& +~\sum\limits_{i\, =
\,1}^2 {h_i \,\left( {e_\mu ^{\left( i \right)} \,n_\nu \,\, - \,\,e_\nu
^{\left( i \right)} \,n_\mu } \right)} +
\nonumber
\\
&&+~e\,\left( {p_\mu \,n_\nu
\,\, - \,\,p_\nu \,n_\mu } \right),
\label{psi1}
\\
\tilde  \psi \,_{\left[ {\mu \,\nu } \right]} \,\left( {\underline{p}}
\right)&=&\tilde f\,\left( {e_\mu ^{\left( 1 \right)} \,e_\nu ^{\left(
2 \right)} \,\, - \,\,e_\nu ^{\left( 1 \right)} \,e_\mu ^{\left( 2 \right)}
} \right)+
\nonumber
\\
&&+~\sum\limits_{i\, = \,1}^2 {\tilde g_i \,\left( {e_\mu
^{\left( i \right)} \,p_\nu \,\, - \,\,e_\nu ^{\left( i \right)} \,p_\mu }
\right)} +
\nonumber
\\
&&+~\sum\limits_{i\, =
\,1}^2 \tilde h_i \,\left( {e_\mu ^{\left( i \right)} \,n_\nu -
e_\nu^{\left( i \right)} \,n_\mu } \right)+
\nonumber
\\
&&+~\tilde e\,\left( {p_\mu \,n_\nu - p_\nu \,n_\mu } \right).
\label{psi2}
\eeqn
\end{subequations}

Further we take into account that the system (\ref{sys4})
is invariant at the gauge transformations
\beqn
\label{eq48}
\delta \,\psi \,_{\left[ {\mu \,\nu } \right]} \,\left( x \right)\,\,\, =
\,\,\,\partial _\mu \,\lambda _\nu \,\left( x \right)\,\, - \,\,\partial
_\nu \,\lambda _\mu \,\left( x \right)+
\nonumber
\\
+~\varepsilon _{\mu \,\nu
\,\alpha \,\beta } \,\partial _\alpha \,\tilde {\lambda }_\beta \,\left( x
\right),
\eeqn
where the gauge finctions $\lambda_\mu (x)$ and $\tilde {\lambda }_\mu (x)$
satisfy the conditions
\beq
\label{BBox1}
\Box\lambda _\mu  -\partial _\mu \partial_\nu \lambda _\nu=0,
\quad
\Box\tilde {\lambda }_\mu -\partial_\mu\partial_\nu \tilde {\lambda }_\nu =0.
\eeq

Keeping in mind the symmetry between the tensors
$\psi \,_{\left[ {\mu \,\nu }
\right]}$ and  $\,\,\,\tilde {\psi }\,_{\left[ {\mu \,\nu } \right]} \,$ one can
replace (\ref{eq48}) by
\begin{equation}
\label{eq49}
\begin{array}{l}
 \delta \,\psi \,_{\left[ {\mu \,\nu } \right]} \,\left( x \right)=
\partial _\mu \,\lambda _\nu \,\left( x \right)\,\, - \,\,\partial
_\nu \,\lambda _\mu \,\left( x \right),
\\
\\
 \delta \,\tilde {\psi }\,_{\left[ {\mu \,\nu } \right]} \,\left( x
\right)=\partial _\mu \,\tilde {\lambda }_\nu \,\left( x
\right)\,\, - \,\,\partial _\nu \,\tilde {\lambda }_\mu \,\left( x
\right),
 \end{array}
\end{equation}
where $\lambda _\mu \,\left( x \right)$ and $\tilde {\lambda }_\mu \,\left(
x \right)$ still satisfy (\ref{BBox1}).
As in the case of the equation (\ref{Dalam}), the
solutions of (\ref{BBox1}) are superpositions of the plane waves
\beqn
\label{eq50}
 \lambda _\mu \,\left( x \right)=\int {\lambda _\mu \left(
{\underline{p}} \right)\,e^{i\,p\,x}\,d^3p}  + h.c.,
\\
 \tilde {\lambda }_\mu \,\left( x \right)=\int {\tilde
{\lambda }_\mu \,\left( {\underline{p}} \right)\,e^{i\,p\,x}\,d^3p}  +h.c.,
\eeqn
with the only difference is that now the amplitudes
$\lambda _\mu \,\left( {\underline{p}}
\right)$ and $\tilde {\lambda }_\mu \,\left( {\underline{p}} \right)$
being expanded over the basis (\ref{basis}) have the structure
\begin{equation}
\label{eq51}
\begin{array}{l}
 \lambda _\mu \,\left( {\underline{p}} \right)\,\,\, =
\,\,\,\sum\limits_{i\, = \,1}^2 {\alpha _i \,e_\mu ^{\left( i \right)} \,\,
+ \,\,\beta \,p_\mu } \,\,, \\
 \tilde {\lambda }_\mu \,\left( {\underline{p}} \right)\,\,\, =
\,\,\,\sum\limits_{i\, = \,1}^2 {\tilde {\alpha }_i \,e_\mu ^{\left( i
\right)} \,\, + \,\,\tilde {\beta }\,p_\mu },
 \end{array}
\end{equation}
\noindent
which does not contain terms with $n_\mu $ (due to terms $\partial _\mu
\,\partial _\nu \,\lambda _\nu \,\left( x \right)$ and $\partial _\mu
\,\partial _\nu \,\tilde {\lambda }_\nu \,\left( x \right)$ in (\ref{BBox1})).
Inserting (\ref{eq45}) and (\ref{eq50})-- (\ref{eq51}) in (\ref{eq49}) we obtain
the following form of the gauge transformations for the potentials
$\psi_{\left[ {\mu \,\nu } \right]}$ and $\tilde{\psi }_{\left[ {\mu\nu } \right]}$
\begin{equation}
\label{eq52}
\begin{array}{l}
 \delta \,\psi \,_{\left[ {\mu \,\nu } \right]} \,\left( {\underline{p}}
\right)\,\,\, = \,\,\,i\,\sum\limits_{i\, = \,1}^2 {\alpha _i \,\left(
{e_\mu ^{\left( i \right)} \,p_\nu \,\, - \,\,e_\nu ^{\left( i \right)}
\,p_\mu } \right)} \,\,, \\
 \delta \,\tilde {\psi }\,_{\left[ {\mu \,\nu } \right]} \,\left(
{\underline{p}} \right)\,\,\, = \,\,\,i\,\sum\limits_{i\, = \,1}^2 {\tilde
{\alpha }_i \,\left( {e_\mu ^{\left( i \right)} \,p_\nu \,\, - \,\,e_\nu
^{\left( i \right)} \,p_\mu } \right)} \,\,. \\
 \end{array}
\end{equation}
The expressions (\ref{eq52}) show that terms containing $g_i$ and $\tilde {g}_i $
in (\ref{psi1}) and (\ref{psi2}) are inessential.
Therefore, one can eliminate them by an appropriate choice of the parameters
$\alpha _i$ and $\tilde {\alpha }_i$ $( \alpha_i =-i\,g_i$ and $
\tilde {\alpha }_i \,\, = \,\, - i\,\tilde {g}_i $), i.e. if one puts
\begin{equation}
\label{eq53}
\psi \,_{\left[ {2\,3} \right]}  =\psi \,_{\left[ {3\,1}
\right]}  = \tilde {\psi }\,_{\left[ {2\,3} \right]}
=\tilde {\psi }\,_{\left[ {3\,1} \right]}  =0.
\end{equation}
Using the definition (\ref{eq43}) for the tensor $\tilde {\psi }\,_{\left[ {\mu
\,\nu } \right]}$ and a relation following from this definition
\begin{equation}
\label{eq54}
\psi \,_{\left[ {\alpha \,\beta } \right]} = -
\frac{1}{2}\,\varepsilon _{\alpha \,\beta \,\mu \,\nu } \,\tilde {\psi
}_{\,\left[ {\mu \,\nu } \right]},
\end{equation}
we obtain from (\ref{eq53})
\begin{equation}
\label{eq55}
\psi \,_{\left[ {1\,4} \right]}=\psi \,_{\left[ {2\,4}
\right]} =\tilde {\psi }\,_{\left[ {1\,4} \right]} =
\tilde {\psi }\,_{\left[ {2\,4} \right]}=0.
\end{equation}
As a result, the decompositions (\ref{psi1}) and (\ref{psi2}) take the form
\begin{equation}
\label{eq56}
\begin{array}{l}
 \psi_{\left[ {\mu \nu } \right]}\left( {\underline{p}}
\right)=f\left( {e_\mu ^{\left( 1 \right)} e_\nu ^{\left(
2 \right)} - e_\nu ^{\left( 1 \right)} e_\mu ^{\left( 2 \right)}
} \right) + e\left( {p_\mu n_\nu -p_\nu n_\mu } \right),
 \\
 \tilde {\psi }_{\left[ {\mu \,\nu } \right]} \left( {\underline{p}}
\right)=\tilde {f}\left( {e_\mu ^{\left( 1 \right)} e_\nu
^{\left( 2 \right)} - e_\nu ^{\left( 1 \right)} e_\mu ^{\left( 2
\right)} } \right) + \tilde {e}\left( {p_\mu n_\nu -p_\nu n_\mu } \right).
 \end{array}
\end{equation}

The expressions (\ref{eq56}) show that the tensor-potential
$\psi \,_{\left[ {\mu
\,\nu } \right]}$ contains only two independent components
either corresponding to the state of the massless spin-1 field
with the longitudinal
polarization.
In the literature, such a field is known as ``notoph'' \cite{ogievetsky66} or
``Kalb-Ramond field'' \cite{kalb74}. Because the system (\ref{sys4})
contains also the potentials  $\psi _0(x)$
and $\tilde {\psi }_0 \,\left( x \right)$, we conclude that
this system (or matrix equation (\ref{eq42}) that is equivalent to it)
describes the Kalb-Ramond field and the massless scalar field with
a doubled set of states degenerated over an additional quantum number.

The massless field systems (\ref{sys3}) and (\ref{sys4}) like the DKE
for a massive particle have an internal symmetry.
Making the use of the matrix form of these systems (\ref{P_1}) and (\ref{eq42})
and explicit form of the matrices $\Gamma _\mu$, $ {\Gamma}'_\mu$, $ P_1 $, and
$P_2 $ one can show that the system symmetry narrows up to the group
$SO({3,1})$ of which generators are determined by the matrices
${\Gamma }'_{\left[ i \right. } {\Gamma}'_{\left. k \right]}$
and $ {\Gamma }'_5 {\Gamma }'_k $.

\section{ Massless ``fermion'' limit}

The DKE is an equation describing a free massive Dirac particle with mass
$m$.
Therefore, a natural way to pass on to the massless limit in (\ref{sys5})
is to put there  $m_1=m_2=0$. We will denote such a transition
as ``fermion'' limit.
As applied to the system of the tensor equations (\ref{sys5}),
this transition leads the systems
\begin{subequations}
\label{sub1}
\beqn
\label{eq57}
\partial _\nu \,\psi \,_{\left[ {\mu \,\nu } \right]} \,\, + \,\,\partial
_\mu \,\psi _0= 0,
\\
\label{eq58}
\frac{1}{2}\,\varepsilon _{\mu \,\nu \,\alpha \,\beta } \,\partial _\nu
\,\psi \,_{\left[ {\alpha \,\beta } \right]} \,\, + \,\,\partial _\mu
\,\tilde {\psi }_0 = 0,
\\
\label{eq59}
\partial _\mu \,\psi _\mu=0,
\\
\label{eq60}
\partial _\mu \,\tilde {\psi }_\mu =0,
\\
 - \partial _\mu \,\psi _\nu \,\, + \,\,\partial _\nu \,\psi _\mu \,\, +
\,\,\varepsilon _{\mu \,\nu \,\alpha \,\beta } \,\partial _\alpha \,\tilde
{\psi }_\beta  =0.
\label{eq61a}
\eeqn
\end{subequations}
This system disintegrates over the Lorenz group into the two
subsystems (\ref{eq57}),(\ref{eq58}) and (\ref{eq59})--(\ref{eq61a}).

It is evident that each of these subsystems can be written down in a matrix form
with field functions $U\left( x \right)$ and $V\left( x \right)$ composed from
$\psi _0 $, $\tilde {\psi }_0$, $\psi \,_{\left[ {\mu \,\nu } \right]} $, and
$\psi _\mu$, $\tilde {\psi }_\mu $, respectively. In addition, if relatively to
the Lorenz group, the function $U(x)$ is transformed as a direct sum $T$ of
presentations for a bivector, a scalar, and a pseudoscalar then an equation for
$U(x)$ is transformed as a direct sum $R$ of
presentations for a vector and a pseudovector. For the function $V(x)$ and
an equation for it the presentations $T$ and $R$ are interchanged.
This means that for each of the subsystems
 (\ref{eq57}),(\ref{eq58}) and (\ref{eq59})--(\ref{eq61a})
taken separately, a massless analog is absent
(a field function for a massive equation and the equation itself are
transformed over the same presentation of the Lorenz group).
Therefore, only bilinear forms like
$\bar {U}\left( x \right)\beta _\mu \,\partial _\mu \,V\left( x \right)$  and
$\bar {V}\left( x \right)\,\beta _\mu \,\partial _\mu \,U\left( x \right)$
are Lorenz-invariant. Here $\beta _\mu $ are 8 by 8 matrices of equations
for $U(x)$ and  $V(x)$. Thus, although the subsystems
(\ref{eq57}),(\ref{eq58}) and (\ref{eq59})--(\ref{eq61a})
are independent algebraically the Lagrangian formulation for them is impossible.
The requirements of the Lagrangian formulation of a theory and existence of its
massive analog lead to the need for combined consideration of
the subsystems (\ref{eq57}),(\ref{eq58}) and (\ref{eq59})--(\ref{eq61a}).
This circumstance is of crucial importance because the energy-momentum density
turns out to be zero for such a massless system.

Indeed, the Lagrangian for the system (\ref{sub1}) can be written as
\beqn
 L&=& -~\psi _\mu \,\partial _\mu \,\psi _0 \,\, -
\,\,\frac{1}{2}\,\psi \,_{\left[ {\mu \,\nu } \right]} \,\left( {\partial
_\mu \psi _\nu \,\, - \,\,\partial _\nu \,\psi _\mu } \right)+
 \nonumber
 \\
&&+~\tilde {\psi }_\mu \,\partial _\mu \,\tilde {\psi }_0 +
\frac{1}{2}\,\varepsilon _{\mu \,\nu \,\alpha \,\beta }
\,\psi \,_{\left[ {\mu \,\nu } \right]} \,\partial _\alpha \,\tilde {\psi
}_\beta .
\label{eq61}
 \eeqn

Inserting (\ref{eq61}) in the expression for the energy-momentum
\begin{equation}
\label{eq62}
T_{\mu \,\nu } \,\,\, = \,\,\,\frac{\partial \,L}{\partial \left(
{\frac{\partial \,\Psi _A }{\partial \,x_\mu }} \right)}\,\frac{\partial
\,\Psi _A }{\partial \,x_\nu } -\delta _{\mu \,\nu } \,L,
\end{equation}
using further the obtained formula and equations (\ref{sub1})
and keeping in mind that terms like full divergency can be omitted,
one finally has
\begin{equation}
\label{eq63}
T_{\mu \,\nu } = 0.
\end{equation}

Hence, the fermion massless limit of the DKE treated as equations describing
a classical boson field (a particle with variable spin 0,1) leads to
zero energy-momentum density.

\section{Gauge-invariant mixing of massless fields}

In papers \cite{kalb74,cremmer74} a non-Higgs mechanism
 to generate masses was proposed.
The method relies on gauge-invariant mixing (the topological interaction)
an electromagnetic field  and a massless
vector field with the zero helicity
(the so-called ${\hat {B}}\wedge{\hat F}$-theory).
The final result of that theory
is the Duffin-Kemmer equation for a massive spin-1 particle.

Such an approach is very important for the string theory
(see \cite{birm91,khou99,spalucci00,bizd07}
and references there in). It is, therefore, important to generalize
this approach for the case of massless systems of the DK type involving
the complete set  of antisymmetric tensor fields in the space with
dimension $d=4$.

At first, we will proceed from
the matrix formulation (\ref{P_1}) and (\ref{eq42})
of the systems (\ref{sys3}) and (\ref{sys4}).
We replace the notation $\Psi$ in (\ref{P_1}) by
 $\Phi$, $\varphi _0$, $\tilde {\varphi }_0$, $\varphi _\mu$,
 $\tilde {\varphi }_\mu$, $\varphi _{\,\left[
{\mu \,\nu } \right]} $. Now the Lagrangian of the matrix equations
(\ref{P_1}) and (\ref{eq42}) takes the form
\begin{equation}
\label{eq64}
L_0 = - \overline \Phi \,\left( {\Gamma _\mu \,\partial _\mu \,\, + \,\,P_1
} \right)\,\Phi \,\, - \,\,\overline \Psi \,\left( {\Gamma _\mu \,\partial
_\mu \,\, + \,\,P_2 } \right)\,\Psi \,\,\,\,,
\end{equation}
\noindent
where $\overline \Phi  = \Phi ^ + \,\Gamma _4 \,{\Gamma }'_4
\,,\,\,\,\,\overline \Psi \,\, = \,\,\Psi ^ + \,\Gamma _4 \,{\Gamma }'_4 $.
We add now to $L_0 $ a term
\begin{equation}
\label{eq65}
L_{int} =- \,m\,\overline \Phi \,P_2 \,\Psi -m\,\overline \Psi \,P_1 \,\Phi ,
\end{equation}
which does not destroy the gauge invariance of the Lagrangian
(\ref{eq64}) relatively to transformations
(\ref{gaugetr}), (\ref{lambdas}), (\ref{eq48}), and (\ref{BBox1}).
One can obtain then the following matrix equations from the total Lagrangian
$L=L_0+L_{int} $
\beqn
\label{eq66}
\Gamma _\mu \,\partial _\mu \,\Phi  + P_1 \,\Phi  + m\,P_2
\,\Psi &=&0,
\\
\label{eq67}
\Gamma _\mu \,\partial _\mu \,\Psi  + P_2 \,\Psi  + m\,P_1\Phi &=&0.
\eeqn
Multiplying (\ref{eq66}) from the left by the matrix $P_1$ and using
relations (\ref{eq13}), we obtain an equation
\begin{equation}
\label{eq68}
\Gamma _\mu \,\partial _\mu \,P_1 \,\Phi + m\,P_2 \,\Psi =0.
\end{equation}
Analogously, multiplying (\ref{eq67}) from the left by the matrix $P_1$
one has
\begin{equation}
\label{eq69}
\Gamma _\mu \,\partial _\mu \,P_2 \,\Psi  + m\,P_1 \,\Phi =0.
\end{equation}
If one puts together (\ref{eq68}) and (\ref{eq69}) and introduces a
notation
\begin{equation}
\label{eq70}
{\Psi }' = P_1 \Phi  + P_2 \,\Psi  =\left(
{\varphi _0 \,,\,\,\,\tilde {\varphi }_0 \,,\,\,\,\psi _\mu \,,\,\,\,\tilde
{\psi }_\mu \,,\,\,\,\varphi _{\left[ {\mu \,\nu } \right]} } \right),
\end{equation}
one arrives at the following equation
\begin{equation}
\label{eq71}
\left( {\Gamma _\mu \partial_\mu + m} \right)\,\Psi'  = 0,
\end{equation}
\noindent
that ut to the notation of the components of the wave function
coincides with the DKE (\ref{2.2}) and (\ref{Psi}).

Therefore, the DKE for massive particles can be represented as a result of
a gauge-invariant mixture of two massless systems, namely dial
symmetric generalizations of electromagnetic field and the Kalb-Ramond field
(notoph). All known in the literature constructions of the analogous mechanism
for tensor fields (an abelian case) of different ranks in the $d=4$ space are
particular cases of the considered approach.

Now let us consider the tensor field system (46) together with another
system of the same kind
\beqn
\label{61a}
\nonumber
\partial _\nu \varphi _{\left[ {\mu \,\nu } \right]} + \partial _\mu \varphi
_0 = 0,
\\
\nonumber
\frac{1}{2}\varepsilon _{\mu \nu \,\alpha \,\beta } \partial _\nu \varphi
_{\left[ {\alpha \,\beta } \right]} + \partial _\mu \tilde {\varphi }_0 =
0,
\\
\partial _\mu \varphi _\mu = 0,
\\
\nonumber
\partial _\mu \tilde {\varphi }_\mu = 0,
\\
\nonumber
 - \partial _\mu \varphi _\nu + \partial _\nu \varphi _\mu + \varepsilon
_{\mu \,\nu \,\alpha \,\beta } \partial _\alpha \tilde {\varphi }_\beta =
0.
\eeqn
The Lagrangian of these systems takes the structure
\begin{equation}
\label{eq80}
L_0 = L_{01} + L_{02} ,
\end{equation}
where Lagrangians $L_{01} $ and $L_{02} $ have the form (\ref{eq61}). The
topological interaction of both systems can be described by the Lagrangian
\beqn
\label{eq81}
L_{int} &=& m~\left( a \varphi _0 \psi _0 + b\tilde {\varphi }_0 \tilde
{\psi }_0 + c\varphi _\mu \psi _\mu + \right.
\nonumber
\\
&&\left.+~d\tilde {\varphi }_\mu \tilde {\psi
}_\mu + e\varphi _{\left[ {\mu \,\nu } \right]} \psi _{\left[ {\mu \,\nu }
\right]}  \right).
\eeqn

From the total Lagrangian $L = L_0 + L_{int} $ one can obtain the following
equations
\beqn
\partial _\mu \varphi _\mu + a m\psi _0 = 0,
\nonumber
\\
\partial _\mu \tilde {\varphi }_\mu - bm\tilde {\psi }_0 = 0,
\nonumber
\\
\partial _\nu \varphi _{\left[ {\mu \,\nu } \right]} + \partial _\mu \varphi
_0 - cm\psi _\mu = 0,
\nonumber
\\
\frac{1}{2}\varepsilon _{\mu \,\nu \,\alpha \,\beta } \partial _\nu \varphi
_{\left[ {\alpha \,\beta } \right]} + \partial _\mu \tilde {\varphi }_0 +
dm\tilde {\psi }_\mu = 0,
\nonumber
\\
 - \partial _\mu \varphi _\nu + \partial _\nu \varphi _\mu + \varepsilon
_{\mu \,\nu \,\alpha \,\beta } \partial _\alpha \tilde {\varphi }_\beta +
2em\psi _{\left[ {\mu \,\nu } \right]} = 0,
\\
\partial _\nu \psi _\mu + a m\varphi _0 = 0,
\nonumber
\\
\partial _\mu \tilde {\psi }_\mu - bm\tilde {\varphi }_0 = 0,
\nonumber
\\
\partial _\nu \psi _{\left[ {\mu \,\nu } \right]} + \partial _\mu \psi _0 -
cm\varphi _\mu = 0,
\nonumber
\\
\frac{1}{2}\varepsilon _{\mu \,\nu \,\alpha \,\beta } \partial _\nu \psi
_{\left[ {\alpha \,\beta } \right]} + \partial _\mu \tilde {\psi }_0 +
dm\tilde {\varphi }_\mu = 0,
\nonumber
\\
 - \partial _\mu \psi _\nu + \partial _\nu \psi _\mu + \varepsilon _{\mu
\,\nu \,\alpha \,\beta } \partial _\alpha \tilde {\psi }_\beta + 2em\varphi
_{\left[ {\mu \,\nu } \right]} = 0.
\nonumber
\eeqn
Then we can make transition from the equations to the two systems
\beqn
\label{62a}
\partial _\mu \Lambda _\mu + am\Lambda _0 = 0,
\nonumber
\\
\partial _\mu \tilde {\Lambda }_\mu - bm\tilde {\Lambda }_0 = 0,
\nonumber
\\
\partial _\nu \Lambda _{\left[ {\mu \,\nu } \right]} + \partial _\mu \Lambda
_0 - cm\Lambda _\mu = 0,
\\
\frac{1}{2}\varepsilon _{\mu \,\nu \,\alpha \,\beta } \partial _\nu \Lambda
_{\left[ {\alpha \,\beta } \right]} + \partial _\mu \tilde {\Lambda }_0 +
dm\tilde {\Lambda }_\mu = 0,
\nonumber
\\
 - \partial _\mu \Lambda _\nu + \partial _\nu \Lambda _\mu + \varepsilon
_{\mu \,\nu \,\alpha \,\beta } \partial _\alpha \tilde {\Lambda }_\beta +
2em\Lambda _{\left[ {\mu \,\nu } \right]} = 0
\nonumber
\eeqn
and
\beqn
\nonumber
\\
\partial _\mu \Omega _\mu - am\Omega _0 = 0,
\nonumber
\\
\partial _\mu \tilde {\Omega }_\mu + bm\tilde {\Omega }_0 = 0,
\nonumber
\\
\partial _\nu \Omega _{\left[ {\mu \,\nu } \right]} + \partial _\mu \Omega
_0 + cm\Omega _\mu = 0,
\\
\frac{1}{2}\varepsilon _{\mu \,\nu \,\alpha \,\beta } \partial _\nu \Omega
_{\left[ {\alpha \,\beta } \right]} + \partial _\mu \tilde {\Omega }_0 -
dm\tilde {\Omega }_\mu = 0,
\nonumber
\\
 - \partial _\mu \Omega _\nu + \partial _\nu \Omega _\mu + \varepsilon _{\mu
\,\nu \,\alpha \,\beta } \partial _\alpha \tilde {\Omega }_\beta - 2em\Omega
_{\left[ {\mu \,\nu } \right]} = 0,
\nonumber
\eeqn
where
\beqn
\label{eq85}
\begin{array}{l}
 \Lambda _0 = \varphi _0 + \psi _0 ,\,\,\,\,\,\tilde {\Lambda }_0 = \tilde
{\varphi }_0 + \tilde {\psi }_0 ,\,\,\,\,\,\,\Lambda _\mu = \varphi _\mu +
\psi _\mu , \\
 \tilde {\Lambda }_\mu = \tilde {\varphi }_\mu + \tilde {\psi }_\mu
,\,\,\,\,\,\,\Lambda _{\left[ {\mu \,\nu } \right]} = \varphi _{\left[ {\mu
\,\nu } \right]} + \psi _{\left[ {\mu \,\nu } \right]}, \\
 \Omega _0 = \varphi _0 - \psi _0 ,\,\,\,\,\,\,\tilde {\Omega }_0 = \tilde
{\varphi }_0 - \tilde {\psi }_0 ,\,\,\,\,\,\,\Omega _\mu = \varphi _\mu -
\psi _\mu , \\
 \tilde {\Omega }_\mu = \tilde {\varphi }_\mu - \tilde {\psi }_\mu
,\,\,\,\,\,\,\Omega _{\left[ {\mu \,\nu } \right]} = \varphi _{\left[ {\mu
\,\nu } \right]} - \psi _{\left[ {\mu \,\nu } \right]} . \\
 \end{array}
\eeqn

In the case of choosing
\begin{equation}
\label{eq86}
a = d = 1,\quad b = c = - 1,\quad e = \frac{1}{2}
\end{equation}
we obtain two types of the DKE.

The matrix form of the Lagrangian of the equations (46) and (\ref{61a}) is
\begin{equation}
\label{eq87}
L_0 = - \bar {\Phi }\Gamma _\mu \partial _\mu \Phi - \bar {\Psi }\Gamma _\mu
\partial _\mu \Psi .
\end{equation}
We add to (\ref{eq87}) the term

\begin{equation}
\label{eq88}
L_{int} = - m\bar {\Phi }\Psi - m\bar {\Psi }\Phi .
\end{equation}
As a result, we obtain the matrix equations
\beqn
\label{eq89}
\Gamma _\mu \partial _\mu \Phi + m\Psi = 0,
\\
\Gamma _\mu \partial _\mu \Psi + m\Phi = 0.
\label{eq90}
\eeqn
Putting together (\ref{eq89}) and (\ref{eq90}) and introducing the notation
\beq
\label{eq91}
\Lambda = \Phi + \Psi ,\,\,\,\,\,\Omega = \Phi - \Psi ,
\eeq
one arrives at the equations
\beqn
\label{eq92}
\left( {\Gamma _\mu \partial _\mu + m} \right)\,\Lambda = 0,
\\
\label{eq93}
\left( {\Gamma _\mu \partial _\mu - m} \right)\,\Omega = 0,
\eeqn
which are the matrix analogs of the tensor systems (\ref{62a})--(\ref{eq86}).

\section{Conclusion}

We have investigated there massless limits of the DKE.
It is shown that first of the limits leads to a two-potential
formulation of the electrodynamics in the Feynman gauge.
Second one gives a generalized description of the Ogievetsky-Polubarinov notoph.
Third limit corresponds to the massless field with zero energy density.
The method to generate masses through the gauge-invariant mixing
(the topological interaction) of massless fields
(${\hat {B}}\wedge{\hat F}$-theory)
 is generalized for the case of above massless systems of the DK type.
As a result, the DKE of two types for particles with masses are found.
One of these equations after quantization can obey  Bose-Einstein statistics
and second one can obey Dirac-Fermi statistics \cite{SS87,PS88}.
This means that from the topological interaction
of the massless DK fields we can proceed to the tensor
and Dirac fields with masses.

\end{document}